\documentstyle[prl,aps,epsf,multicol]{revtex}
\begin{document}
\draft
\title{Generic
entanglement generation, quantum statistics and complementarity}
\author{S. Bose $^1$ and D. Home $^2$}

\address{$^1$ Centre for Quantum Computation, Clarendon Laboratory,
    University of Oxford,
    Parks Road,
    Oxford OX1 3PU, England}
\address{$^2$ Physics Department, Bose Institute, Calcutta 700009, India}

\maketitle
\begin{abstract}
A general and an arbitrarily efficient scheme for entangling the
spins (or any spin-like degree of freedom) of two independent
uncorrelated identical particles by a combination of two particle
interferometry and which way detection is formulated. It is shown
that the same setup could be used to identify the quantum
statistics of the incident particles from either the sign or the
magnitude of measured spin correlations. Our setup also exhibits a
curious complementarity between particle distinguishability and
the amount of generated entanglement.
\end{abstract}


\begin{multicols}{2}
 Recent years have witnessed a great surge of interest in the
applications of entanglement
\cite{bennett00,zeilinger97,braunstein98}. In this context, it is
important to explore efficient and general ways of preparing
entanglement. Most known mechanisms for obtaining entangled
states \cite{braunstein98,zeilinger95,keiss93,haroche97,gheri},
are dependent on the {\em specific nature} of the systems
involved. Here we propose a very {\em general} scheme for
entangling the spins (or any spin-like degree of freedom) of two
particles of {\em any} type (bosons or fermions) by a combination
of two particle interferometry and which way detection. The
fractional yield of entangled pairs for a given number of input
pairs can be arbitrarily increased by a recursive procedure using
just one beam splitter and two detectors. The main application of
our setup will be in entangling material objects such as neutrons,
electrons, atoms or macromolecules. This will enable testing
quantum nonlocality through separate measurements on  far
separated massive particles. A salient feature of our setup is
the fact that two independent identical particles {\em do not
need to interact directly} for getting entangled. They only need
to interact individually with beam splitters and detectors and
their {\em indistinguishability} can be exploited to yield
entanglement. This is thus useful for entangling those particles
which interact weakly (or do not interact) with other particles
of the same species. Another advantage of our setup is that the
disentangled initial state of the independent particles can be
prepared  by classical communications without any non-linearity
as opposed to the states in the current methods
\cite{braunstein98,zeilinger95}. We show that the same setup
allows {\em identification of the quantum statistics} of the
incident particles from either the presence/absence of
entanglement or the sign of measured spin correlations, depending
upon whether the two-particle input state is unpolarized or
oppositely spin polarized, as opposed to previous tests based on
particle number measurements
\cite{hanbury,mandel87,zeilinger97b,yamamoto98,loudon98}. It also
exhibits a complementarity between particle distinguishability
and the amount of entanglement produced. This complementarity
involving "which particle" information in two particle
interference, differs fundamentally from the usual form involving
"which way" information in single particle interference
\cite{scully91,rempe98,buks98,englert96}. Our work also suggests a
curious {\em dual} to standard entanglement in the context of
second quantization.

%\vspace*{-0.2cm}

\begin{figure}
\begin{center}
\leavevmode \epsfxsize=6cm \epsfbox{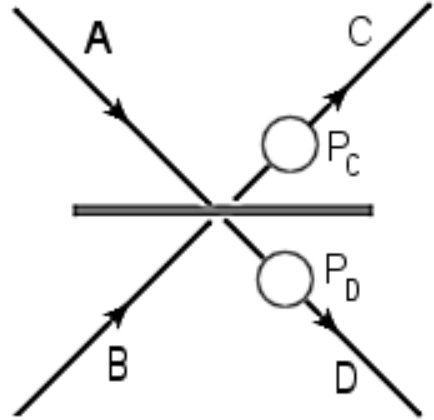}
 \caption{\narrowtext A preliminary setup
consisting of a beam splitter (input paths A and B, output paths
C and D) and absorptionless path detectors $P_C$ and $P_D$ which
do not disturb the spin. When a pair of identical particles with
opposite spins are incident on the first beam splitter, one from
arm $A$ and the other from arm $B$, then corresponding to a
coincidence in $P_C$ and $P_D$, a spin entangled state is
generated.}
\label{setup}
\end{center}
\end{figure}

 We  will first present a preliminary setup ({\em not} our {\em
 complete
 scheme})
 by modifying an interference process
 used in the last stage of production of polarization entangled of photons through
 down-conversion \cite{zeilinger95,mandel87b} which
 produces
entangled states with $50$ percent efficiency (successful cases
being identifiable by appropriate detector clicks). Fig.1 depicts
the setup composed of a beam splitter with input channels $A$ and
$B$, output channels $C$ and $D$ and which-channel detectors
$P_C$ in $C$ and $P_D$ in $D$. These detectors are assumed to be
{\em nonabsorbing} and are able to {\em determine the path without
disturbing the spin} (this is possible since position and spin
commute; feasibility discussed later). Now consider two identical
particles in different spin states (say $|\uparrow\rangle$ and
$|\downarrow\rangle$) incident simultaneously on the beam
splitter from arms $A$ and $B$ as shown in Fig.1. This state, in
second quantized notation, is described as $a^{\dagger}_{A
\uparrow} a^{\dagger}_{B \downarrow} |0\rangle$ where $|0\rangle$
is the vacuum state and $a^{\dagger}_{A \uparrow}$ and
$a^{\dagger}_{B \downarrow}$ are creation operators for
$\uparrow$ spin in path $A$ and $\downarrow$ spin in path $B$
respectively. We will label the state concisely as $|A \uparrow;
B \downarrow \rangle$. For fermions, $|A \uparrow; B \downarrow
\rangle = -|B \downarrow; A \uparrow \rangle$ and for bosons $|A
\uparrow; B \downarrow \rangle = |B \downarrow; A \uparrow
\rangle$.  The transformation done by the beam splitter is
\cite{loudon98,stenholm99}

\begin{eqnarray}
|A \uparrow; B \downarrow \rangle  &\rightarrow&
\frac{1}{\sqrt{2}}\{\frac{1}{\sqrt{2}}(|D \uparrow; C \downarrow
\rangle \pm |D \downarrow ; C \uparrow  \rangle)\} \nonumber \\&+&
\frac{i}{2}(| C \uparrow ; C \downarrow  \rangle + |D \uparrow ;D
\downarrow \rangle),
\label{onebs}
\end{eqnarray}
where the $+$ sign stands for fermions and the $-$ sign stands for
bosons. After the detectors click, the combined state of the
particles and the detectors is
\begin{eqnarray}
\frac{1}{\sqrt{2}}\{\frac{1}{\sqrt{2}}(|D \uparrow; C \downarrow
\rangle &\pm& |D \downarrow ; C \uparrow  \rangle)\}
|P_C^{*}\rangle |P_D^{*}\rangle \nonumber \\\oplus \frac{i}{2}(| C
\uparrow ; C \downarrow  \rangle |P_C^{*}\rangle
|P_D\rangle&\oplus& |D \uparrow ;D \downarrow \rangle |P_C\rangle
|P_D^{*}\rangle),
\end{eqnarray}
where $\{|P_C\rangle,|P_D\rangle\}$ and
$\{|P_C^{*}\rangle,|P_D^{*}\rangle\}$ are the unexcited and
excited (corresponding to detection of one or more particles)
detector states respectively. In the above $\oplus$ has been used
to indicate the lack of {\em coherence} between orthogonal
detector states. When the detectors are found in the state
$|P_C^{*}\rangle |P_D^{*}\rangle$ (coincidence), the state of the
particles is projected onto $\frac{1}{\sqrt{2}}(|D \uparrow; C
\downarrow \rangle \pm |D \downarrow ; C \uparrow \rangle)$. The
spin part of this state can be rewritten in the first quantized
notation (using the paths as particle labels) as
$|\psi^{\pm}\rangle_{CD}=\frac{1}{\sqrt{2}}(|\uparrow\rangle_D|
\downarrow \rangle_C \pm | \downarrow \rangle_D | \uparrow
\rangle_C)$ (spin entangled state). It is fully legitimate to use
the paths as particle labels because the particles are {\em
identical} (the same labeling is used for photon pairs exiting a
parametric down converter \cite{zeilinger95}).

In the above description, we have made two significant changes to
the scheme used for photons. First is the presence of the special
detectors, which could be easier to design for massive particles.
For photons \cite{zeilinger95,mandel87b}, such detectors are
absent and {\em after} the path measurements, the resource of
entanglement is {\em not available} for applications. Our
detectors help us to obtain an useful {\em source} of spin
entangled particles. The other difference with the photonic scheme
is that the incident state $|A \uparrow; B \downarrow \rangle$, is
{\em not} entangled. In contrast, the down-conversion based
schemes \cite{zeilinger95} use a {\em momentum entangled}
incident state of the two photons. We merely require the
particles to impinge on the beam splitter at the same instant of
time; any {\em prior entanglement} is {\em not} necessary for
this. For example, if atoms from two independent sources are
velocity selected, they can be made to pass through the same
region at the same time as in recent cavity QED experiments
\cite{haroche01}. By removing the necessity of non-linearity for
generating the incident state, we thus enlarge the scope of the
method to cover all types of particles.

\begin{figure}
\begin{center}
\leavevmode \epsfxsize=6cm \epsfbox{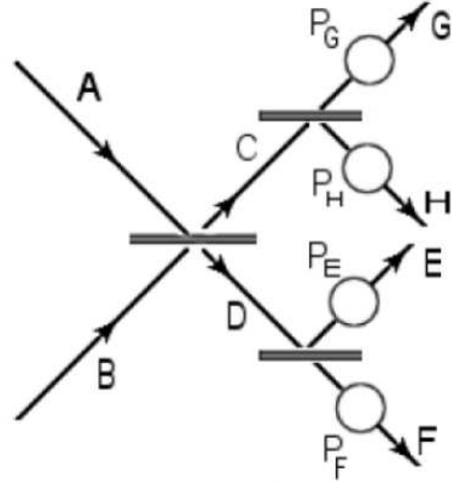} %\vspace*{3cm}
\caption{\narrowtext An improved version of the setup. A and B are
the input paths, E,F,G and H are the output paths and
$P_E,P_F,P_G$ and $P_H$ are absorptionless path detectors which
do not disturb the spin. A pair of identical particles with
opposite spins are incident on the first beam splitter, one from
arm $A$ and the other from arm $B$. For coincidence between any
pair of detectors, which happens in $75$ percent of the cases, a
spin entangled state is emitted along the corresponding pair of
paths}
\label{setup2}
\end{center}
\end{figure}

 We now describe our full scheme, in which the efficiency can be
arbitrarily increased. First we consider the addition of two more
beam splitters to the setup as shown in Fig.2 with the four exit
paths $E, F, G$ and $H$ being incorporated with which-path
detectors $P_E, P_F, P_G$ and $P_H$ respectively. If the state $|A
\uparrow; B \downarrow \rangle$ is incident on the first beam
splitter, then the final combined state of the particles and the
detectors is
\begin{eqnarray}
\ominus(|G \uparrow; E \downarrow \rangle \pm |G \downarrow ; E
\uparrow \rangle)|P_E^{*}\rangle |P_F\rangle |P_G^{*}\rangle
|P_H\rangle &\nonumber \\\oplus  i(|H \uparrow; E \downarrow
\rangle \pm |H \downarrow ; E \uparrow \rangle)|P_E^{*}\rangle
|P_F\rangle |P_G\rangle |P_H^{*}\rangle & \nonumber \\\oplus i(|G
\uparrow; F \downarrow \rangle \pm |G \downarrow ; F \uparrow
\rangle)|P_E\rangle |P_F^{*}\rangle |P_G^{*}\rangle |P_H\rangle &
\nonumber
\\\oplus (|H \uparrow; F \downarrow \rangle \pm
|H \downarrow ; F \uparrow \rangle) |P_E\rangle |P_F^{*}\rangle
|P_G\rangle |P_H^{*}\rangle &\nonumber
\\\ominus (|F \uparrow; E \downarrow \rangle \mp
|F \downarrow ; E \uparrow \rangle) |P_E^{*}\rangle
|P_F^{*}\rangle |P_G\rangle |P_H\rangle\nonumber &
\\\ominus (|H \uparrow; G \downarrow \rangle \mp |H
\downarrow ; G\uparrow \rangle)|P_E\rangle |P_F\rangle
|P_G^{*}\rangle |P_H^{*}\rangle &\nonumber \\ \ominus  i\{|E
\uparrow; E \downarrow \rangle |P_E^{*}\rangle |P_G\rangle
\ominus |G \uparrow; G \downarrow \rangle |P_E\rangle
|P_G^{*}\rangle \}&|P_F\rangle |P_H\rangle\nonumber
\\\oplus  i \{ |F \uparrow; F \downarrow \rangle
|P_F^{*}\rangle |P_H\rangle \oplus   |H \uparrow; H \downarrow
\rangle |P_F\rangle |P_H^{*}\rangle \}&|P_E\rangle |P_G\rangle
\nonumber
\end{eqnarray}
where the upper/lower signs stand for fermions/bosons and
$\oplus$ and $\ominus$ indicate the lack of coherence. The above
expression indicates that there will be coincidence between a
pair of detectors in $75$ percent of the cases. In each of these
cases, a spin entangled state will be generated along the
corresponding pair of exit channels. For example, for fermions,
if $P_E$ and $P_G$ click,  $|\psi^{+}\rangle_{GE}$ is produced
and if $P_H$ and $P_G$ click, $|\psi^{-}\rangle_{GH}$ is
produced. Based on the knowledge of the detector clicks, all the
different entangled states can be converted to a desired
entangled state by applying spin dependent phases along
appropriate paths. Note that, only in $25$ percent of the cases
detector clicks will result in a disentangled state.

    The above improvement in success probability stems from the fact that
the extra pair of beam splitters not only map the entangled part
$|\psi^{\pm}\rangle_{DC}$ of the state after the first beam
splitter to entangled final parts, but also map $50$ percent of
the disentangled parts \cite{keiss93} $| C \uparrow ; C
\downarrow \rangle$ and $| D \uparrow ; D \downarrow \rangle$ to
entangled final parts . We can easily double the number of output
channels by subdividing each of the existing outputs by beam
splitters and at each stage the entangled fraction increases. If
we subdivide in this manner to obtain $2^N$ outputs, and put
detectors only in these final exit paths, the fractional yield of
entangled pairs is $1-1/{2^N}$. For $N=7$, this exceeds $99$
percent.

  We now describe an interesting {\em
  feedback scheme} to reduce the resources ({\em i.e}
  the number of beam splitters) required in our proposal, while
  still retaining the arbitrarily high efficiency. In the setup of Fig.\ref{setup}),
  when the output is
  found disentangled (i.e, $P_C$ and $P_D$ do not click in coincidence),
  it is fed back again into the {\em same} beam splitter. This procedure could be repeated successively to
increase arbitrarily the efficiency of entanglement generation.
  Here, the probability of failure decreases exponentially ($2^{-N}$) with the
  number $N$ of feedback rounds, while the required
resources (a single beam splitter and a pair of detectors) remain
unchanged.

 The next issue of the paper is the identification of quantum
statistics through spin correlation measurements. Consider Fig.1
once again and the incident state $|A \uparrow; B \downarrow
\rangle$. If there is a detector coincidence, then the unitary
operation $|\uparrow\rangle \rightarrow
\frac{1}{\sqrt{2}}(|\uparrow\rangle +|\downarrow\rangle),
|\downarrow\rangle \rightarrow \frac{1}{\sqrt{2}}
(|\uparrow\rangle -|\downarrow\rangle)$ is applied to the spins
of particles in each of the output channels. The spins are then
measured in the $\{|\uparrow\rangle,|\downarrow\rangle\}$ basis.
There will be {\em perfect correlation} between the spin
measurement outcomes in the two paths for {\em fermions} and {\em
perfect anticorrelation} for {\em bosons}. The sign of the spin
correlation can thus be used to identify quantum statistics of
the incident particles. This differs from earlier schemes which
rely on particle number measurements for testing statistics
\cite{hanbury,mandel87,zeilinger97b,yamamoto98,loudon98}.

 In our setup (Fig.\ref{setup}), if each particle is fed in the
 random spin state $\rho_{M}=|\uparrow\rangle\langle\uparrow|+|\downarrow\rangle\langle\downarrow|$,
 the difference between bosons and fermions is
 manifested through the {\em presence} or
 {\em absence} of entanglement in a given experiment. For
 detector coincidence, a maximally entangled state $|\psi^{-}\rangle_{CD}$ is created for
 bosons and
 a disentangled state $(1/2)|\uparrow_C \uparrow_D \rangle\langle
\uparrow_C\uparrow_D|
+(1/2)|\downarrow_C\downarrow_D\rangle\langle
\downarrow_C\downarrow_D|+ (1/\sqrt{2})|\psi^{-
}\rangle_{CD}\langle \psi^{-}|_{CD}$ results for fermions. The
difference between bosons and fermions can thus be demonstrated
through Bell's inequalities. The sufficiency of $\rho_{M}$
implies that no prior spin correlation between the incident
particles is necessary for entanglement generation using bosons.

   We now describe a curious complementarity
between particle distinguishability and entanglement in our
scheme. The complementarity of ``which channel" information with
fringe contrast in single particle interference is much discussed
\cite{scully91,rempe98,buks98,englert96}. In two particle
interferometry, ``which channel" information is naturally
replaced by ``which particle" information. The particles impinging
on our setup (Fig.1) are indistinguishable apart from their spins
(which we choose not to measure as we intend to create a spin
entangled state). Now suppose, the particles were partially or
fully distinguishable through some other observable such as
energy or momentum or any non-spin internal degree of freedom.
For example, suppose the incident state is $|A \uparrow S_1; B
\downarrow S_2\rangle$, with $|\langle S_1|S_2 \rangle|=a \leq
1$. Then the two-particle state produced due to detector
coincidence is $\frac{1}{\sqrt{2}}(|D \uparrow S_1; C \downarrow
S_2\rangle \pm |D \downarrow S_2; C \uparrow S_1\rangle)$. The
spin state of the particles (in the first quantized notation) is
\begin{eqnarray}
\rho=\frac{1}{2}&(&|\uparrow_C \downarrow_D \rangle\langle
\uparrow_C\downarrow_D| +|\downarrow_C\uparrow_D\rangle\langle
\downarrow_C\uparrow_D|\nonumber\\&\pm&|a|^2 |\uparrow_C
\downarrow_D \rangle\langle \downarrow_C\uparrow_D|\pm|a|^2
|\downarrow_C\uparrow_D\rangle \langle \uparrow_C\downarrow_D|)
\end{eqnarray}
Note that in the second quantized notation all the degrees of
freedom belong to the {\em same} Hilbert space (are created from
the same vacuum). But they become elements of {\em distinct}
Hilbert spaces when we proceed to the first quantized notation.
For the above state, a certain entanglement measure called
concurrence \cite{wootters98} is ${\bf E}=|a|^2$. The probability
of successful discrimination between the states $|S_1 \rangle$ and
$|S_2 \rangle$ (which is a measure of particle
distinguishability) is ${\bf D}=1-|a|^2$. Thus we have, in
analogy with Englert's relation in single particle interference
\cite{englert96}, the following testable complementarity relation
\begin{equation}
\label{compl}
 {\bf E}+{\bf D}=1.
\end{equation}
The concurrence ${\bf E}$ for $\rho$ can be inferred by measuring
the expectation value of the Bell-CHSH operator
$\hat{a}\hat{b}+\hat{a}\hat{b^{'}}+\hat{a^{'}}\hat{b}-\hat{a^{'}}\hat{b^{'}}$
on the two particles (labeled by their paths $C$ and $D$) with
$\hat{a}=\sigma^{C}_x, \hat{a^{'}}=\sigma^{C}_y,
\hat{b}=\frac{1}{\sqrt{2}}(\sigma^{D}_x+\sigma^{D}_y),\hat{b^{'}}=\frac{1}{\sqrt{2}}(\sigma^{D}_x-\sigma^{D}_y)$
and dividing the result by $\pm 2\sqrt{2}$.

  Eq.(\ref{compl}) also helps us to estimate the amount of entanglement
generated by our scheme if $|S_1 \rangle$ and $|S_2 \rangle$ are
Gaussian wave-packet states of the incident particles arriving at
the beam splitter. If the wave packets have width $\sigma$,
velocity $v$, and a time delay $\Delta t$ with respect to each
other, the entanglement is {\bf E}$=\exp(-v^2\Delta
t^2/2\sigma^2)$.

   We now suggest an
an unexplored ramification of entanglement in the context of
second quantization invoked in our treatment. One needs {\em at
least two
   quantum numbers}
   associated with the creation operators (the spin and output channel labels in our case) to
   meaningfully describe an entangled state. In a suitable quantum state such as
   $\frac{1}{\sqrt{2}}(|D
\uparrow; C \downarrow \rangle \pm |D \downarrow ; C \uparrow
\rangle)$, when one of the labels (the
   channel labels $C$ and $D$ in our case) denotes the identity of
   the particle, the other degree of freedom (the spin in our
   case) appears {\em entangled}. However, the above state can
   {\em equally well} be rewritten in first quantized notation as
   $|D_{\uparrow} C_{\downarrow}\rangle + |C_{\uparrow}
   D_{\downarrow}\rangle$ (for both bosons and fermions), where the spins have been used
   to label the particles and their paths are entangled. Devising
   schemes to detect this {\em dual form of entanglement} (via the {\em reversal} of
   labels) is
   currently under way \cite{bose-home}.

    Finally we discuss the feasibility of our scheme. The three
basic ingredients are: beam splitters, the capability of
performing two particle interference, and which way detectors
that keep the internal degree of freedom to be entangled
undisturbed. Beam splitters are available for photons, electrons
\cite{tonomura98}, neutrons \cite{rauch91} , atoms
\cite{rempe98,stenholm99,schmiedmayer} and even for macromolecules
\cite{zeilinger99}. Two particle interferometry is feasible with
photons \cite{mandel87,zeilinger97b}. Its realizability with
electrons has received much attention of late
\cite{yamamoto98,tonomura98}. For atoms, one could use the
recently fabricated beam splitters for guided atoms
\cite{schmiedmayer} to study two particle interferometry, as
discussed in Ref.\cite{stenholm99}. Next is the question of the
special type of "which way" detectors required for our scheme. A
theoretical model of an absorptionless path detectors that keep
spin unaffected has been considered in the context of quantum
state reduction \cite{scully78}. Such detectors have already been
fabricated for electrons (based on the effects of electric fields)
for a complementarity experiment \cite{buks98} and there are also
proposals for simple variants of such detectors \cite{blencowe}.
Detectors of the required type have been suggested for photons
(based on crossed phase modulation) \cite{grangier91}, and for
neutrons (based on momentum transfer) \cite{movmir}. For atoms,
one can  implement our scheme by placing cavities in the arms $C$
and $D$ to act as our which way detectors. One would have to use
an atom with hyperfine ground levels
$|g_1\rangle,|g_2\rangle,|g_3\rangle$ and $|g_4\rangle$ which can
be made to interact with a cavity field in Fock state $|n\rangle$
to undergo transitions $|g_1\rangle |n\rangle\rightarrow
|g_2\rangle|n+1\rangle$ and $|g_3\rangle|n\rangle \rightarrow
|g_4\rangle|n+1\rangle$ \cite{kimble95}. Then, with an incident
state $|A g_1; B g_3 \rangle$, appropriate transitions in
cavities will result in the entangled state
$\frac{1}{\sqrt{2}}(|g_2\rangle_D| g_4 \rangle_C \pm | g_4
\rangle_D | g_2 \rangle_C)$ when both cavities are found in the
state $|n+1\rangle$. For macromolecules with many internal
degrees of freedom, one can choose any two independent degrees of
freedom, one for entangling and the other for path detection.

  To summarize, we have presented a scheme of arbitrarily high efficiency for
entangling two particles of {\em any type}. This is important, as
entangled states of objects such as neutrons, electrons or
macromolecules are yet to be prepared. Our scheme provides strong
motivation for developing two particle interferometry for various
systems in tandem with absorptionless "which-way" detectors. That
the {\em same} setup can be used to test quantum statistics
through entanglement-induced spin correlations and probe
complementarity in two particle interference enhance the
significance of our scheme. Our work also suggests potential
connections between entanglement, quantum statistics and
complementarity, which call for further study (See for example,
Ref.\cite{omar01}) .

We thank D. Atkinson  and D. Bouwmeester for useful discussions.
The feedback scheme was stimulated by the comments of R. Srikanth
and a referee. DH acknowledges the Royal Society - Indian
National Science Academy Exchange Programme which helped to
initiate this collaboration. We also acknowledge the First SERC
Winter School on Foundations of Quantum Mechanics and Quantum
Optics (Department of Science and Technology, Govt. of India).

%\begin{multicols}

\end{multicols}


\begin{references}

\bibitem{bennett00}
C. H. Bennett and D. P. DiVincenzo, Nature {\bf 404}, 247 (2000).

%

\bibitem{zeilinger97}
D. Bouwmeester {\em et. al.}, Nature {\bf 390}, 575 (1997); D.
Boschi {\em et. al.}, Phys. Rev. Lett. {\bf 80}, 1121 (1998).

\bibitem{braunstein98}
 A. Furusawa {\em et. al.}, Science {\bf 282}, 706 (1998).
%
%
\bibitem{zeilinger95}
P. G. Kwiat {\em et. al.}, Phys. Rev. Lett. {\bf 75}, 4337 (1995).
%
\bibitem{keiss93}
T. E. Kiess {\em et. al.}, Phys. Rev. Lett. {\bf 71}, 3893 (1993).

%
\bibitem{haroche97}
E. Hagley {\em et. al.}, Phys. Rev. Lett. {\bf 79}, 1 (1997); C.
A. Sackett {\em et. al.}, Nature {\bf 404}, 256 (2000).

%

\bibitem{gheri}
C. Saavedra {\em et. al.}, Phys. Rev. A {\bf 61}, 062311 (2000);
A. Kuzmich and E. S. Polzik,  Phys. Rev. Lett. {\bf 85}, 5639
(2000); S. Bose {\em et. al.},  Phys. Rev. Lett. {\bf 83}, 5158
(1999); A. Beige, W. J. Munro and P. L. Knight, Phys. Rev. A {\bf
62} 2102, (2000).

%
\bibitem{hanbury}
R. Hanbury-Brown and R. Q. Twiss, Nature (London) {\bf 27}, 177
(1956).
%
\bibitem{mandel87}
C. K. Hong, Z. Y. Ou and L. Mandel, Phys. Rev. Lett. {\bf 59}
2044 (1987).


%
\bibitem{zeilinger97b}
G. Weihs {\em et. al.}, Phys. Rev. A {\bf 54}, 893 (1996).
%
\bibitem{yamamoto98}
R. C. Liu {\em et. al.}, Nature {\bf 391}, 263 (1998).
%
%
\bibitem{loudon98}
R. Loudon, Phys. Rev. A {\bf 58}, 4904 (1998).
%

%
\bibitem{scully91}
M. O. Scully, B.-G. Englert and H. Walther, Nature {\bf 351},
111-114 (1991); B.-G. Englert, M. O. Scully and H. Walther, Am. J.
Phys. {\bf 67}, 325(1999).

\bibitem{rempe98}
S. Durr, T. Nonn and G. Rempe,  Nature {\bf 395}, 33 (1998).
%
\bibitem{buks98}
E. Buks {\em et. al.}, Nature {\bf 391}, 871-874 (1998).
%
\bibitem{englert96}
 B.-G.
Englert, Phys. Rev. Lett. {\bf 77}, 2154 (1996).
%
\bibitem{mandel87b}
Z. Y. Ou, C. K. Hong and L. Mandel, Opt. Comm. {\bf 63}, 118
(1987); Z. Y. Ou and L. Mandel, Phys. Rev. Lett. 61, 50 (1988)M.
Pavi$\breve{c}$i$\acute{c}$,  J. Opt. Soc. Am. B {\bf 12}, 821
(1995).
%
\bibitem{stenholm99}
E. Andersson, M. T. Fontenelle and S. Stenholm,   Phys. Rev. A
{\bf 59} 3841 (1999).

%
\bibitem{haroche01}
S. Osnaghi {\em et. al.}, Phys. Rev. Lett. {\bf 87}, 037902
(2001).

\bibitem{wootters98}
W.~K.~Wootters, Phys. Rev. Lett. {\bf 80}, 2245 (1998).

\bibitem{bose-home} S. Bose and D. Home, (in preparation).

%
\bibitem{tonomura98}
T. Kodama {\em et. al.},
  Phys. Rev. A {\bf 57}, 2781 (1998).




%

%



\bibitem{rauch91}
S. A. Werner {\em et. al.}, Phys. Rev. Lett. {\bf 67}, 683 (1991).


%
\bibitem{schmiedmayer}
D. Cassettari {\em et. al.}, Phys. Rev. Lett. {\bf 85}, 5483
(2000).









%
\bibitem{zeilinger99}
M. Arndt {\em et. al.}, Nature {\bf 401}, 680 (1999).

\bibitem{scully78}
M. O. Scully, R. Shea and J. D. McCullen,  Phys. Repts. {\bf 43},
485 (1978).

%
\bibitem{blencowe}
A. D. Armour and M. P. Blencowe, Phys. Rev. B {\bf 64}, 035311
(2001).











%
\bibitem{grangier91}
P. Grangier and J.-F. Roch, Opt. Comm. {\bf 83}, 269 (1991).


%
\bibitem{movmir}
D. M. Greenberger and A. Yasin, Found. Phys. {\bf 19}, 679 (1989).





\bibitem{kimble95}
A. S. Parkins {\em et. al.}, Phys. Rev. A {\bf 51}, 1578-1596
(1995).




\bibitem{omar01}
Y. Omar {\em et. al.}, quant-ph/0105120.









\end{references}
\end{document}